\documentclass[twocolumn,showpacs,aps,prl,preprintnumbers,amsmath,amssymb]{revtex4}

\usepackage{graphicx}
\usepackage{dcolumn}
\usepackage{bm}
\usepackage{epsfig}

\begin{document}
\date{\today}


\title{Backscattering of Dirac fermions in HgTe quantum wells with a finite gap}

\author{G. Tkachov, C. Thienel, V. Pinneker, B. B\"uttner, C. Br\"une, H. Buhmann, L. W. Molenkamp, and E. M. Hankiewicz}

\affiliation{
Fakult\"at f\"ur Physik und Astronomie and R\"ontgen Center for Complex Material Systems, Universit\"at W\"urzburg, Am Hubland, 97074 W\"urzburg, Germany}

\begin{abstract}
The density-dependent mobility of n-type HgTe quantum wells with inverted band ordering has been studied both experimentally and theoretically. While semiconductor heterostructures with a parabolic dispersion exhibit an increase in mobility with carrier density, high quality HgTe quantum wells exhibit a distinct mobility maximum. We show that this mobility anomaly is due to backscattering of Dirac fermions from random fluctuations of the band gap (Dirac mass).
Our findings open new avenues for the study of Dirac fermion transport with finite and random mass, which so far has been hard to access.
\end{abstract}
\maketitle

{\em Introduction}.--
Topological insulators, such as HgTe quantum wells (QWs)~\cite{Bernevig06,Koenig07Full,Roth09Full},
Bi$_{1-x}$Sb$_x$~\cite{Fu07,Hsieh08Full}, Bi$_2$Se$_3$~\cite{Xia09Full}, and Bi$_2$Te$_3$~\cite{Chen09Full,Zhang09}
provide the most recent examples of the realization of Dirac fermions in condensed matter physics.
Unlike graphene~\cite{Neto09}, where two valleys of Dirac fermions exist,
in these new materials Dirac fermions appear only at a single point in the Brillouin zone.
The absence of valley scattering and the possibility of
unconventional types of disorder, such as a random Dirac mass~\cite{Ludwig94},
make electron transport studies on these materials particularly interesting.
However, the low carrier mobility of most topological insulators still poses a serious obstacle
for transport investigations of specific scattering mechanisms.
The notable exception are MBE-grown HgTe QWs where transport measurements have been used to detect
the quantum spin Hall state~\cite{Koenig07Full} and, more recently,
to realize massless single-valley two-dimensional Dirac fermions~\cite{Buettner10Full}.

In this paper, we study both experimentally and theoretically a new manifestation of Dirac fermion transport
in n-type HgTe QWs, i.e., the occurrence of a maximum in the mobility as a function of carrier density. The origin of the maximum is the competition of two disorder scattering mechanisms, viz. scattering by charged impurities and by QW width fluctuations which induce a fluctuating band gap, or, equivalently, fluctuating Dirac mass.
As in other semiconductor heterostructures \cite{Shur87}, in HgTe QWs the screening of ionized impurities by the carriers results, initially, in a monotonic increase of the mobility with increasing carrier density.
However, while the impurity scattering becomes weaker with increasing carrier density,
the other scattering mechanism - well-width fluctuations - becomes increasingly important, leading to a reduction of the carrier mobility. Dirac mass disorder generates scattering between states of opposite momenta, also called backscattering. Thus, the observed mobility peak is a clear manifestation of Dirac fermion backscattering in HgTe QWs.

Backscattering of Dirac fermions is most pronounced in gapped systems~\cite{Ando05}. 
As a consequence, the mobility of graphene does not show a maximum, but rather a saturation at high carrier densities (see, e.g. Refs.~\cite{Neto09,Ando05,Nomura06,Chen08}), which has been attributed to charged impurity scattering. Although some theoretical studies indicate that mass disorder may play a role in the vicinity of the neutrality point of graphene (see, e.g. Refs.~\cite{Ziegler09,Bardarson10}), its experimental identification has remained problematic.

\begin{figure}[b!]
\includegraphics[width=80mm]{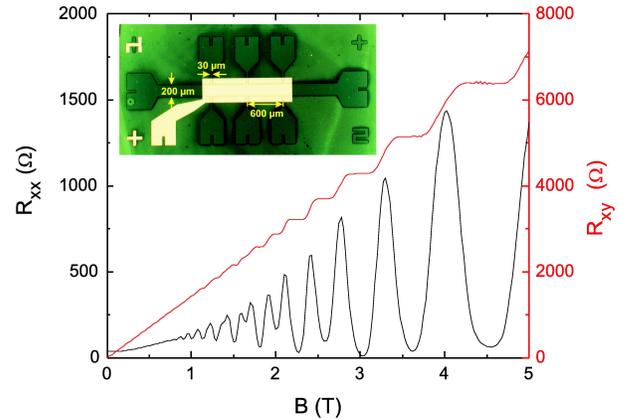}
\caption{
Hall and magnetoresistance for sample \#6 at zero gate voltage. The measurement yields an electron mobility of
$1.1 \times 10^6$ cm$^{-2}$/Vs and a carrier density of $4.3 \times 10^{11}$ cm$^{-2}$. Inset: Micrograph of the Hallbar structure.
}
\label{h}
\end{figure}

{\em Experiment}.--
Transport experiments have been performed on modulation doped HgTe/Hg$_{0.3}$Cd$_{0.7}$Te QW structures 
fabricated by molecular beam epitaxy on lattice-matched (Cd,Zn)Te substrates. 
The samples have been patterned into Hall bar devices with dimensions of $(600 \times 200)$ $\mu$m$^2$ 
using a low temperature optical lithography process, and covered by a 5/100 nm Ti/Au gate electrode 
which is deposited onto a 110 nm thick Si$_3$N$_4$/SiO$_2$ multilayer gate insulator. Ohmic contacts are provided by thermal indium bonding. A micrograph of the  structure is shown in the inset of Fig.~\ref{h}. The samples have nominal QW widths, $d$, ranging from 5.0 to 12.0 nm, thus covering both the normal 
($d < 6.3$ nm) and inverted ($d > 6.3$ nm) band structure regimes \cite{Koenig07Full,Buettner10Full}. 
The relevant parameters of all samples are summarized in Table~\ref{Table}. 
We have performed standard Hall and Shubnikov-de Haas measurements on these samples in magnetic fields up to $B = 5$~T, at a temperature 4.2~K. 
As an example, Hall and magnetoresistance data for sample \#6, which has the peak highest mobility are shown in Fig.~\ref{h}. 
The carrier densities and mobility of ungated samples ($V_g = 0$) are in the range of $3.0 \times 10^{11}$ cm$^{-2}$ to $5.5 \times 10^{11}$ cm$^{-2}$ 
and several $10^5$ cm$^{2}$/Vs, respectively.

\begin{figure}[t]
\includegraphics[width=80mm]{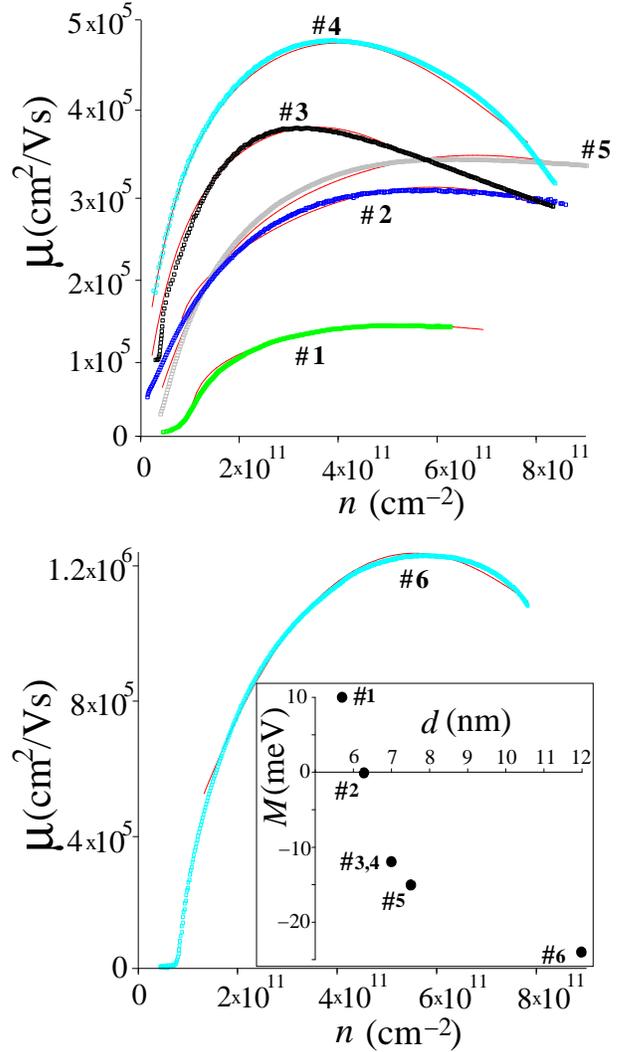}
\caption{
Mobility $\mu$ versus carrier density $n$ for six HgTe QWs ($\# 1-\# 6$).
The experimental $\mu(n)$ dependence (thick lines), including the appearance of the maximum,
agrees well with our model (thin red lines) that takes into account
the competition of charged-impurity disorder and QW thickness fluctuations.
Inset: Band gap $M$ versus QW thickness $d$ [sample $\# 1$ has a positive band gap, 
sample $\# 2$ has approximately zero gap, while the rest $\# 3-\# 6$ are in the inverted regime].
The pronounced maximum is observed in high quality inverted samples $\# 3$, $\# 4$ and $\# 6$
where the impurity concentration is sufficiently low (see also Table~\ref{Table}).
}
\label{Mu}
\end{figure}

The density dependent carrier mobilities are obtained from the gate-voltage dependence of the longitudinal resistance at zero magnetic field, 
while the dependence of the carrier concentration on gate-voltage was deduced from the Hall voltage measured at a fixed magnetic field of 300 mT. 
The thick solid lines in Fig.~\ref{Mu} show the experimentally observed mobility versus carrier density for the various HgTe QWs. 
In high quality samples with inverted subband structure ordering  (\#3, \#4, and \#6) 
the mobility exhibits a distinct maximum for carrier densities in the range of 3 to  6 $\times 10^{11}$ cm$^{-2}$. 
For samples with a slightly lower mobility (\#1, \#2, and \#5) a saturation of the mobility with carrier density is observed.

{\em Model}.--
In order to explain the unusual dependence of the mobility on carrier density we build a model for the conductivity in HgTe QWs which is based on the four-band Dirac model of Refs.~\cite{Bernevig06,Koenig07Full}. The effective Dirac Hamiltonian has the following form:
\begin{eqnarray}
\hat{H}=s_z\mbox{\boldmath$\sigma$}( A {\bf\hat k} + M_{\bf\hat k}{\bf z} ) + D{\bf\hat k}^2 - E_{_F}
+ V_{\bf r} + \delta M_{\bf r} s_z\sigma_z,
\label{H}
\end{eqnarray}
where $M_{\bf\hat k}=M(d) + B{\bf\hat k}^2$.
In Eq.~(\ref{H}) the linear term, which is proportional to ${\bf\hat k}=-i\nabla_{\bf r}$ and the constant $A$ originate from the hybridization of the first electron (E1) and heavy-hole (HH1) subbands in the quantum well. These two subbands are represented by the pseudospin $\mbox{\boldmath$\sigma$}$,
whose components $\sigma_x$, $\sigma_y$ and $\sigma_z$ are $2 \times 2$ Pauli matrices.
(The real spin degree of freedom is represented by the Pauli matrix $s_z$.)
The term proportional to the effective Dirac mass $M_{\bf\hat k}$ reflects the average band gap, $|M_{\bf\hat k=0}|=|M(d)|$, which is determined by the nominal thickness $d$ of the QW.
The ${\bf k}$-dependent part of the mass (the $B{\bf\hat k}^2$ -term) and the parabolic background $D{\bf\hat k}^2$ take further details of the band dispersion in HgTe QWs into account~\cite{Bernevig06,Koenig07Full}.
Note that the mass term violates the pseudo-time reversal symmetry ${\bf\hat k}\to -{\bf\hat k}$ and
$\mbox{\boldmath$\sigma$}\to -\mbox{\boldmath$\sigma$}$ of Hamiltonian (\ref{H}),
which manifests itself in a dependence of the conductivity on $M_{\bf\hat k}$.
Equation (\ref{H}) also takes the two most relevant types of disorder into account:
a random potential due to charged impurities, $V_{\bf r}$,
and spatial fluctuations of the Dirac mass, $\delta M_{\bf r}$, which are
related to deviations of the QW thickness from its average value $d$.
We assume the n-type transport regime, with the Fermi level $E_{_F}$ in the conduction band, 
under the weak scattering condition, $k_{_F}\upsilon_{_F}\tau\gg 1$,
where $\tau$ is the transport relaxation time and $\upsilon_{_F}$ and $k_{_F}=\sqrt{2\pi n}$
are the Fermi velocity and wave-vector, respectively.
The conductivity can then be obtained from the Kubo formula:
\begin{eqnarray}
\sigma_{xx}=\frac{e^2\hbar}{2\pi}
\int\frac{d^2{\bf k}}{(2\pi)^2}
{\rm Tr}
\left[
\hat{v}_x\, \hat{G}^{^R }_{\bf k} \, \hat{\tilde v}_x \, \hat{G}^{^A }_{\bf k}
\right]= e^2 N_{_F} \upsilon^2_{_F} \tau,
\label{Kubo}
\end{eqnarray}
where $\hat{G}^{^{R/A} }_{{\bf k},\epsilon}$ are disorder-averaged retarded and
advanced Green's functions and $\hat{v}_x$, $\hat{\tilde v}_x$ are current vertices
(the tilde indicates vertex renormalization by disorder
in the ladder approximation~\cite{Rammer}). The resulting conductivity $\sigma_{xx}$ is proportional to
the density of states per spin at the Fermi level, $N_{_F}=k_{_F}/2\pi\hbar\upsilon_{_F}$,
and the transport time $\tau=1/\int_0^{2\pi}(1-\cos\theta)w(\theta)d\theta$, 
where $w(\theta)$ is the scattering rate at angle $\theta$. 
Below, we obtain an expression for $w(\theta)$ from the electron self-energy in the self-consistent Born approximation (SCBA), calculate $\tau$ and the carrier-density dependent mobility $\mu(n)=\sigma_{xx}(n)/ne$.

For this purpose, we assume that the potential and mass disorder are
uncorrelated and completely characterized
by the two-point correlation functions
$\langle V_{\bf r}V_{\bf r^\prime} \rangle=\zeta^V_{\bf r-r^\prime}$
and  $\langle \delta M_{\bf r}\delta M_{\bf r^\prime} \rangle=\zeta^M_{\bf r-r^\prime}$.
In ${\bf k}$ space this leads to the Dyson equation
$\hat{G}^{^R}_{\bf k}=\hat{G}^{^R}_{0\bf k} +\hat{G}^{^R}_{0\bf k}\hat{\Sigma}^{^R}_{\bf k}\hat{G}^{^R}_{\bf k}$ for the disorder-averaged Green's function $\hat{G}^{^R}_{\bf k}$ where the self-energy $\hat{\Sigma}^{^R}_{\bf k}$ 
is given by the standard SCBA expression  
\begin{eqnarray}
&&
\hat{\Sigma}^{^R}_{\bf k} =\int ( \zeta^V_{\bf k-q}\hat{G}^{^R}_{\bf q} +
                             \zeta^M_{\bf k-q}s_z\sigma_z\hat{G}^{^R}_{\bf q}s_z\sigma_z  ) \frac{ d{\bf q} }{(2\pi)^2},
\label{Dyson_Sigma}\\
%
&&
\hat{G}^{^R}_{0\bf k}=\frac{1}{2}\frac{I + s_z\mbox{\boldmath$\sigma$}\cdot{\bf e}_{\bf k} }{\epsilon - \xi_{\bf k } },
\quad
{\bf e}_{\bf k}=\frac{  A {\bf k} + M_{\bf k } {\bf z} }{ \sqrt{  A^2{\bf k}^2 +  M^2_{\bf k }  }  }.
\label{G_0}
\end{eqnarray}
The unperturbed Green's function $\hat{G}^{^R}_{0\bf k}$ describes a conduction-band electron
with dispersion $\xi_{\bf k }=\sqrt{  A^2{\bf k }^2 +  M^2_{\bf k }} + D{\bf k }^2 - E_{_F}$
and chirality $s_z\mbox{\boldmath$\sigma$}\cdot{\bf e}_{\bf k}=1$ ($I$ is the unit matrix). 
The solution~\cite{Rammer} for the Green's function 
$\hat{G}^{^R}_{\bf k}=\frac{1}{2}(I + s_z\mbox{\boldmath$\sigma$}\cdot{\bf e}_{\bf k})/
( \epsilon - \xi_{\bf k } + \frac{i\hbar}{2\tau_{el}} )$ contains the finite elastic life-time 
$
\tau_{el}=1/\int_0^{2\pi} w(\theta)d\theta,
$
where the scattering rate $w(\theta)$ at angle $\theta$ is given by 
\begin{eqnarray}
&&
w(\theta)=\frac{ N_{_F} }{\hbar}
\biggl[
\left(
\zeta^V_{ 2k_{_F}|\sin\theta/2| } +
\zeta^M_{ 2k_{_F}|\sin\theta/2| } {\rm e}^2_\perp
\right) \cos^2\frac{\theta}{2}
\nonumber\\
&&
+
\left(\zeta^V_{ 2k_{_F}|\sin\theta/2| } {\rm e}^2_\perp  +
      \zeta^M_{ 2k_{_F}|\sin\theta/2| }
\right) \sin^2\frac{\theta}{2}
\biggr].
\label{w}
\end{eqnarray}
Here ${\rm e}_\perp=M_{k_F}/\sqrt{  A^2k^2_{_F} +  M^2_{k_F} } $
is the out-of-plane component of the unit vector ${\bf \rm e}_{\bf k}$ at $|{\bf k}|=k_{_F}$
[see Eq.~(\ref{G_0})].

\begin{figure}[t!]
\includegraphics[width=80mm]{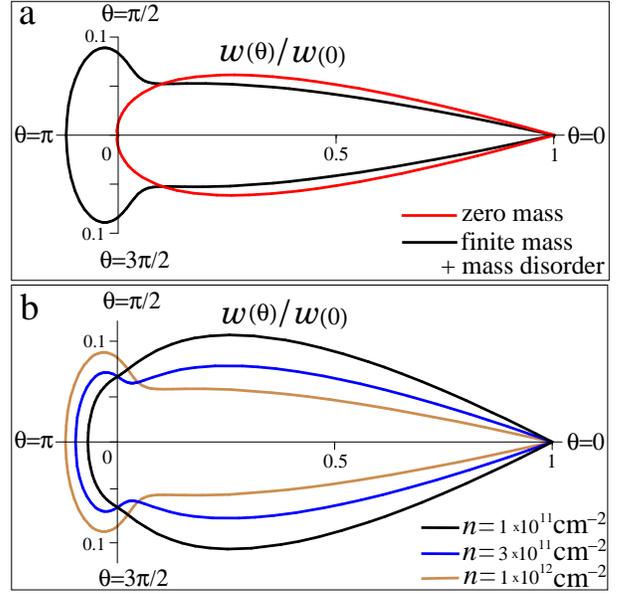}
\caption{
(a) Polar plot of the angular dependence of the normalized scattering rate $w(\theta)$ 
for massless and massive ($M=-15$ meV and $\zeta^M=1.5\times 10^{-3}$) Dirac fermions 
[see Eqs.~(\ref{w}) and (\ref{Corr})].
Massive carriers show strong backscattering at angles $\pi/2 < \theta < 3\pi/2$.
(b) Polar plot of the normalized function $w(\theta)$ 
for different carrier densities $n$ in the inverted regime
($M=-15$ meV and $\zeta^M=1.5\times 10^{-3}$). The large-angle scattering is enhanced with increasing $n$,
which accounts for the non-monotonic mobility $\mu(n)$ in Fig.~\ref{Mu}.
}
\label{w_fig}
\end{figure}

Analyzing Eq.~(\ref{w}), one notes that the first term vanishes at $\theta=\pi$. This is the
well-known absence of backscattering in the limit of massless Dirac electrons~\cite{Ando05}
(this behavior is plotted as the red curve in Fig.~\ref{w_fig}a).
In gapless Dirac materials, the pseudospin $\mbox{\boldmath$\sigma$}$ points along ${\bf k}$, and, therefore, states with  ${\bf k}$ and  $-{\bf k}$ are orthogonal to each other, i.e. unavailable for scattering.
The most essential distinction between HgTe quantum wells and the zero gap case is the second term in Eq.~(\ref{w}), which actually has a maximum at the backscattering angle $\theta=\pi$.
This term originates from the finite Dirac mass $M_{\bf k} s_z\sigma_z$ and its spatial fluctuation
$\delta M_{\bf r} s_z\sigma_z$.  Both of these result in an out-of-plane pseudospin component $\sigma_z$,
hence the opposite ${\bf k}$ states are no longer orthogonal and
large-angle scattering now becomes possible (cf. the black curve in Fig.~\ref{w_fig}a).
Figure~\ref{w_fig}b shows that the backscattering is enhanced with increasing carrier density $n$,
which accounts for the non-monotonic behavior of $\mu(n)$ in inverted quantum wells observed in Fig.~\ref{Mu}.

To proceed further, we make specific assumptions for the correlation functions:
\begin{eqnarray}
 \zeta^V_{\bf k}=n_i ( 2\pi e^2 )^2/(\varepsilon |{\bf k}| + 4\pi e^2 N_{_F} )^2,
 \quad
 \zeta^M_{\bf k}=A^2\zeta^M,
 \label{Corr}
\end{eqnarray}
where $\zeta^V_{\bf k}$ is the usual correlation function of screened Coulomb impurities~\cite{Ando05,Nomura06} 
with the concentration $n_i$ and the average dielectric constant, $\varepsilon\approx 15$, of the HgTe/CdTe QW.
$\zeta^M_{\bf k}$ is normalized such that $\zeta^M$ is a small dimensionless parameter,
which guarantees that $\delta M$ is small compared with the leading linear term $\sim A k_{_F}$
in Hamiltonian (\ref{H}). Since $\delta M$ is caused by fluctuations of the QW thickness,
$\zeta^M$ is independent of carrier density.
As shown below, this approximation yields an excellent agreement with the measurements.
Using Eqs.~(\ref{Kubo}), (\ref{w}) and (\ref{Corr}) we find the mobility $\mu(n)=\sigma_{xx}/ne$: 
\begin{eqnarray}
\mu(n)&=&\frac{4e/h}
{
n_i [\zeta^V_1(n) + \zeta^V_2(n){\rm e}^2_\perp(n)] + 
n \zeta^M \frac{ [3+{\rm e}^2_\perp(n)] A^2}{ \hbar^2\upsilon^2_{_F}(n) } 
},\,\,\,\,\,
\label{Mu_result}
\end{eqnarray}
where
\begin{eqnarray}
&&
{\rm e}_\perp(n) = (M + 2\pi B n)/\sqrt{  2\pi A^2 n +  ( M + 2\pi B n )^2  },
\label{e_result}\\
%
&&
\zeta^V_1=\int_0^\pi d\theta \sin^2\theta
\left[ 1 +\varepsilon \hbar\upsilon_{_F}(n)e^{-2}\sin\theta/2 \right]^{-2},
\label{V1}\\
%
&&
\zeta^V_2=4\int_0^\pi d\theta \sin^4\theta/2 
\left[ 1 + \varepsilon\hbar\upsilon_{_F}(n)e^{-2}\sin\theta/2 \right]^{-2}.\,\,\,\,\,\,\,\,\,\,
\label{V2}
\end{eqnarray}
Using Eq.~(\ref{Mu_result}) we can quantitatively reproduce all experimental curves for $\mu(n)$ in Fig.~\ref{Mu} using the disorder parameters $n_i$, $\zeta^M$ and band gap $M$ indicated in Table~\ref{Table}. The values of $M$ obtained from this fit agree well with those obtained from band structure calculations and the analysis of the experimental SdH oscillations.

\begin{table}[t!]
\caption{Sample and disorder parameters (the other parameters of the model are fixed:
$\varepsilon=15$, $A=0.38$ eV$\cdot$nm, $B=1.2$ eV$\cdot$nm$^2$ and $D=0.85$ eV$\cdot$nm$^2$). }
\centering
\begin{tabular}{ c c c c c c c c }
 \hline\hline
                                       & \# 1  & \# 2   & \# 3    & \# 4    & \# 5    & \# 6 \\ [0.5ex]
 \hline
nominal well width (nm)                & 5.7    & 6.3    & 7      &  7      & 7.5     & 12   \\
max. mobility
($10^5$ cm$^2$/Vs)                   & 1.32  & 2.96   & 3.71    &  4.76   & 3.33    & 12.27   \\
  $M $ (meV)                           & 10    & 0      & -12     & -12     & -15     & -24   \\
  $n_i (\times 10^{10}{\rm cm}^{-2}$)  & 8     & 3.41   & 2.95    &  2.55   & 4.56    & 1.09  \\
  $\zeta^M (\times 10^{-3})$           & 1.65  & 0.5    & 1.65    &  0.91   & 0.43    & 0.33  \\ [1ex]
  \hline
\end{tabular}
\label{Table}
\end{table}

{\em Discussion}.-- 
Using the theoretical model presented in the previous section, we can now explain the observed non-monotonic dependence of $\mu(n)$ as follows. The initial increase in $\mu(n)$ results from the impurity screening: it reflects the density-dependence of the Fermi velocity $\upsilon_{_F}(n)$ which enters the screened impurity potential via the DOS in Eq.~(\ref{Corr}) [see also Eqs.~(\ref{V1}) and (\ref{V2})]. 
Similar behavior was found in conventional doped heterostructures \cite{Shur87}. However, here $\upsilon_{_F}(n)$ is specific to the massive Dirac Hamiltonian (\ref{H}).   
Furthermore, for the inverted quantum wells $\# 3-\# 6$, the mobility is additionally enhanced due to the reduction of the total Dirac gap $-|M(d)| + 2\pi B n$ at low carrier densities $n$ [cf. Eq.~(\ref{e_result})]. 
This leads to a more rapid initial increase in $\mu(n)$ compared to sample $\# 1$, which has a normal band structure, and the zero-gap sample $\# 2$. At higher carrier densities the mobility starts to decrease for all the inverted samples,
most pronouncedly so for the high-quality samples $\# 3$, $\# 4$ and $\# 6$. 
Since the estimated impurity concentration is lowest for these samples (see Table~\ref{Table}), 
we attribute this decrease to the fluctuations of the QW width (Dirac mass),  
accounted for by the term $\propto n\zeta^M$ in Eq.~(\ref{Mu_result}).
Intuitively, the reduction of the mobility can be explained by the fact that the rate of scattering off the well-width fluctuations grows proportionally to the carrier DOS $N_{_F}(n)$ because more states become available for backscattering as the Fermi surface size increases with $n$ [see Eqs. (\ref{w}) and (\ref{Corr})].

From the fits we estimate the amplitude of the well width fluctuations, $\delta d$, to be of the order of $0.2 - 0.3$ nm, 
which is obtained by integrating the correlation function of the thickness fluctuations,
$\langle \delta d_{\bf r}\delta d_{\bf r^\prime} \rangle=\langle \delta M_{\bf r}\delta M_{\bf r^\prime} \rangle/M^\prime(d)^2$, over the area. This integral is a measure of the typical height $\delta d$ times length $L$ of the fluctuation:
$L \delta d \sim A \sqrt{\zeta^M}/|M^\prime(d)|\approx 0.7$ nm$^2$, where
we take $A=0.38$ eV$\cdot$nm and $\zeta^M=10^{-3}$ from Table~\ref{Table} and the gap derivative
$|M^\prime(d)|\approx 17$ meV$\cdot$nm$^{-1}$ from the inset in Fig.~\ref{Mu}.
For a realistic sample $L\gg \delta d$, thus we estimated the ratio $L/\delta d \approx 10$,
which yields $\delta d\approx 0.26$ nm. The result is in good agreement with X-ray reflectivity data 
on MBE-grown HgTe QW structures of similar quality~\cite{Stahl10}.

{\em Conclusions}.--
We have shown both experimentally and theoretically that the density-dependent mobility of high quality HgTe quantum wells with an inverted band structure exhibits a maximum. While the initial increase in the mobility is mainly due to scattering at screened charged impurities, the decreasing part is associated with the band gap fluctuations that generate mass disorder for Dirac-like fermions in this material. Our findings thus clearly demonstrate the occurrence of Dirac
fermion backscattering in finite gap systems.

{\em Acknowledgments}.-- We thank S.-C. Zhang, B. Trauzettel and P. Recher for helpful discussions.
The work was supported by DFG grants HA5893/1-1, SPP 1285/2, and the joint DFG/JST research programme on 'Topological Electronics'.


\begin{thebibliography}{99}

\bibitem{Bernevig06}
B. A. Bernevig and T. L. Hughes and S. C. Zhang,
Science {\bf 314}, 1757 (2006).

\bibitem{Koenig07Full}
M. K{\"o}nig, S. Wiedmann, C. Br{\"u}ne, A. Roth, H. Buhmann, L. W. Molenkamp, X.-L. Qi and S.-C. Zhang,
Science {\bf 318}, 766 (2007).

\bibitem{Roth09Full}
A. Roth, C. Br\"une, H. Buhmann, L. W. Molenkamp, J. Maciejko, X.-L. Qi, and S.-C. Zhang,
Science {\bf 325}, 294 (2009).

\bibitem{Fu07}
L. Fu and C. L. Kane, Phys. Rev. B {\bf 76}, 045302 (2007).  

\bibitem{Hsieh08Full}
D. Hsieh, D. Qian, L. Wray, Y. Xia, Y. S. Hor, R. J. Cava, and  M. Z. Hasan,
Nature Phys. {\bf 452}, 970–974 (2008).

\bibitem{Xia09Full}
Y. Xia, D. Qian, D. Hsieh, L. Wray, A. Pal, H. Lin, A. Bansil, D. Grauer, Y. S. Hor, R. J. Cava, and  M. Z. Hasan,
Nature Phys. {\bf 5}, 398 (2009).

\bibitem{Chen09Full}
Y. L. Chen, J. G. Analytis, J.-H. Chu, Z. K. Liu, S.-K. Mo, X. L. Qi, H. J. Zhang, D. H. Lu, X. Dai, Z. Fang, S. C. Zhang, I. R. Fisher, Z. Hussain, and  Z.-X. Shen,
Science {\bf 325}, 178 (2009).

\bibitem{Zhang09}
H. Zhang, C.-X. Liu, X.-L. Qi, X. Dai, Z. Fang  and  S.-C. Zhang, Nature Phys. {\bf 5}, 438 (2009). 

\bibitem{Neto09}
A.H. Castro Neto, F. Guinea, N.M. Peres, K.S. Novoselov, and A.K. Geim,
Rev. Mod. Phys. {\bf 81}, 109 (2009) and references therein.

\bibitem{Ludwig94}
A. W. W. Ludwig, M. P. A. Fisher, R. Shankar, and G. Grinstein, Phys. Rev. B {\bf 50}, 7526 (1994).

\bibitem{Buettner10Full}
B. B\"uttner, C. X. Liu, G. Tkachov, E. G. Novik, C. Br\"une, H. Buhmann, E. M. Hankiewicz, P. Recher, B. Trauzettel, S. C. Zhang and L. W. Molenkamp, 
Nature Phys. (published online 6 February 2011, doi:10.1038/nphys1914); e-print arXiv:1009.2248.

\bibitem{Shur87}
M. S. Shur, J. K. Abrokwah, R. R. Daniels, and D. K. Arch, J. Appl. Phys {\bf 61}, 1643 (1987).

\bibitem{Ando05}
T. Ando, J. Phys. Soc. Jpn. {\bf 74}, 777 (2005).

\bibitem{Nomura06}
K. Nomura and A.H. MacDonald, Phys. Rev. Lett. {\bf 96}, 256602 (2006).

\bibitem{Chen08}
J.-H. Chen, C. Jang, S. Adam, M. S. Fuhrer, E. D. Williams and M. Ishigami, Nature Phys. {\bf 4}, 377 (2008).

\bibitem{Ziegler09}
K. Ziegler, Phys. Rev. Lett. {\bf 102}, 126802  (2009).

\bibitem{Bardarson10}
J. H. Bardarson, M. V. Medvedyeva, J. Tworzydlo, A. R. Akhmerov, and C. W. J. Beenakker,
Phys. Rev. B {\bf 81}, 121414(R) (2010).

\bibitem{Rammer}
J. Rammer, {\em Quantum Transport Theory} (Westview Press, 2004).
 
\bibitem{Stahl10}
A. Stahl, Ph. D. Thesis, University of W\"urzburg, 2010.

\end{thebibliography}
\end{document}